# Evolutionarily primitive social entities

Angelica Kaufmann

University of Milan

angelica.kaufmann@unimi.it

Social entities only exist in virtue of collective acceptance or recognition, or acknowledgement by two or more individuals in the context of joint activities. Joint activities are made possible by the coordination of plans for action, and the coordination of plans for action is made possible by the capacity for collective intentionality. This paper investigates how primitive is the capacity that nonhuman animals have to create social entities, by individuating how primitive is the capacity for collective intentionality. I present a novel argument for the evolutionary primitiveness of social entities, by showing that the collective intentions upon which these social entities are created and shared are metaphysically reducible to the relevant individual intentions.

KEYWORDS: social entities; collective intentionality; intention; action planning; mental content.

**1. Introduction**

The European Community, the Venice International Film Festival, the Vatican, the NBA, and the Berlin Philharmonic Orchestra are all social entities. And so, this paper argues, are the coalitions that get established during the hunting practices among the chimpanzee populations of the Taï Forest, in the Ivory Coast (Boesch & Boesch-Achermann, 2000; Kaufmann, 2020). These chimpanzees, like other primates, behave in a distinctively sophisticated way: they act jointly. But do these coalitions possess the characteristics necessary for being a social entity? The question at issue in this paper is: can nonhuman animals create social entities?

Scholars in social ontology have debated over the primitiveness of collective intentionality. Searle (1995, 2010) argued that certain features of the capacity to establish social entities are primitive in two ways: first, collective intentionality is evolutionarily primitive, that is, animal species other than humans create and share—at least rudimentary—social entities; secondly, collective intentionality is metaphysically primitive, that is, these social entities are created and shared upon collective intentions which are irreducible to individual distal intentions (Searle, 1990).

Researchers in developmental and comparative psychology found Searle's notion of collective intention captivating (Rakoczy & Tomasello, 2007; Tomasello, 2014). And by embracing this notion, they advocate that the capacity to share intention, and then to create social entities, is uniquely human.

In this paper I argue that Searle's two senses of primitive should be analysed separately. I defend his argument for the primitiveness of the capacity for collective intentionality in the first sense but not in the second sense: I argue that (1) collective intentionality exists among nonhuman animals but (2) it is constructed and shared upon collective intentions which are reducible to individual intentions (Bratman, 2014). I also argue that the reduction of collective intentions to individual intentions is necessary for nonhuman animals to be able to create social entities.

This novel proposal is articulated both on conceptual and on empirical grounds drawing on Michael Bratman's "augmented individualist" account of shared agency,[1] and on Christophe Boesch's cognitive ethology research on chimpanzee group behaviour respectively. Throughout the paper I describe the advantages of framing Boesch's findings within Bratman's account. In addition, recent studies support the view offered in this paper that chimpanzees have a significant understanding of each other actions and plans (Buttelman et al., 2017; Duguid and Melis, 2020; Heesen et al., 2021a; Melis and Tomasello, 2019).

This paper is structured as follows. First, I present Searle's account of collective intentionality, and, secondly, I discuss how, Rakoczy and Tomasello (2007) critically consider Searle's view as an over-ascription of certain cognitive faculties to nonhuman creatures. Even if this is not overtly stated in their paper, I infer that the over-ascription of cognitive faculties that they are criticising concerns the capacity for conceptual thinking that is needed to create and share collective intentions.[2] However, Searle (2010) argues again for the primitiveness of the capacity for collective intentionality. I shall explain why their criticism that Searle is over-ascribing certain cognitive capacities to nonhuman animals is a legitimate objection, but it is unconvincingly discussed, and a similar stance reiterated in more recent works by Tomasello (2016, 2020). For this reason, thirdly, I am going to explain how Searle's argument for the primitiveness of collective intentionality can be articulated differently: arguably, (1) language and conceptual thought are interdependent, and since non-linguistic animals are capable of thought, it follows that non-linguistic animals can entertain mental states with a non-conceptual content; (2) since the sort of mental states that are involved in those joint activities that instantiate collective intentionality can be non-conceptual in content, then non-linguistic animals are capable of entertaining mental states that can establish social entities. I shall call this the *Capacity-based Dependency View (CBDV)*.

---

[1] In Bratman (2014) it is explained that a shared intention is an interpersonal structure of related individual intentions, that jointly serve to coordinate action planning.

[2] The confirmation that my interpretation of their criticism is correct and up to date can be found in Tomasello (2014), especially at pp. 46, 47, 136, 138, 140.

Thus, collective intentionality is evolutionarily primitive: social entities exist among nonhuman animals, but they are constructed and shared upon collective intentions, which are reducible to individual intentions. The latter can be created and shared by creatures that lack the mastery of linguistic capacities. I shall introduce a new—primitive—social entity: call this a *hunting team*. I'm going to analyse Taï Chimpanzees' group hunting (Boesch & Boesch-Achermann, 2000) in the light of this concept. By relying on the Planning Theory (Bratman, 2014), I shall explain that the phenomenon of group hunting, and presumably the joint activities of other primates (as recent evidence shows and I will discuss) can be understood as a case of ascription of individual intentions with meshing sub-plans that instantiate a social entity.

Field work on Taï chimpanzee group behaviour brings robust evidence to explore Searle's intuition that collective intentionality exists in different animal species and, as mentioned earlier, more recent evidence on the mutual understanding of actions and plans among chimpanzees supports the view presented in the paper. However, Searle's reductionist account of the prerequisites for collective intentionality is cognitively too demanding to explain what regulates the social world of nonlinguistic creatures. By explaining that chimpanzees mentally represent the distal intentions of one another in nonconceptual terms we can offer a parsimonious explanation of the cognitive requirements for collective intentionality.

**2. The primitiveness of collective intentionality: analysis and critique**

The study of social entities is called social ontology. According to Searle (1995), social entities only exist in virtue of collective acceptance or recognition, or acknowledgement by two or more individuals in the context of joint activities. Joint activities are made possible by the coordination of plans for action, and the coordination of plans for action is made possible by the creation and ascription of collective intention.[3] Insofar as joint action is the result of the coordination and planning of purposeful actions by two or more individuals towards a shared goal, it is philosophically relevant to individuate under which circumstances a group of individuals performing a joint action, and thereby exercising collective intentionality, can be classified as instantiating a social entity.[4]

According to Searle, social entities exist in virtue of collective acceptance, or recognition or acknowledgment (Searle, 2006, p. 13), and their existence depends on us, they are 'observer dependent', because they would have never existed if there had never been any conscious agents. He thinks collective recognition is necessary, but it is still not sufficient for that entity to be social. That is because social entities are such in virtue of the fact that they are created by a set of 'observer independent' (i.e., objective) thoughts that belong to different individuals. In his view, the minds of individuals create social 'observer dependent' (i.e., subjective) entities through the coordinated planning activity of their 'observer independent' thoughts.

Searle outlines the three primitives that constitute social (and, ultimately, institutional) reality: (1) collective intentionality, (2) the assignment of functions, and (3) the assignment of status functions. He argues that (1) "Human beings along with a lot of other social animals, have the capacity for collective intentionality" (Searle, 2010, p. 43) and that (2) "Humans, and some animals, have the capacity to assign functions to object[5], where the objects do not have the function intrinsically but only in virtue of the collective assignment" (Searle, 2006, p. 17); Searle maintains that what separates humans from other animal species is (3) the capacity of assigning status functions, i.e. assigning constitutive rules and procedures.[6] For him, this is because status functions establish the passage from social reality to institutional reality, and one of the prerequisites for the creation of institutional facts is language, which is a faculty that nonhuman animals lack (for an exhaustive discussion on the role played by language in the creation of institutional reality, see Searle, 2006, pp. 18-21).

---

[3] In the literature on philosophy of mind, the notion of intention is commonly referred to in relation to one of the various "dual-intention theories" (Pacherie, 2008). Searle (1983) distinguishes between prior intentions and intentions in action, Brand (1984) between prospective and immediate intentions, Bratman (1987) between future-directed and present-directed intentions, Mele (1992) between distal and proximal intentions. The present paper deals only with distal intention, individual or collective may these be defined.

[4] Note that not all joint actions are instances of social entities, but all instances of social entities are joint actions.

[5] Social *objects* fall under the more neutral notion of *entities* that I deploy.

[6] Wilson (2007) has an argument for the ascription of this third capacity to nonhuman animals.

Part of the goal of this paper is to disentangle the dispute on the primitiveness of collective intentionality by addressing where we should place the evolutionary origins of this phenomenon, prior to the creation of institutional reality. I do not analyse further the third primitive, i.e., the assignment of status functions. However, at the end of the paper I sketch a proposal for prospective investigations about the sufficient conditions for the emergence of the third primitive, and therefore of institutional entities.

The main concern of this paper is with the first primitive, collective intentionality, in that—according to Searle—this is the basic and sufficient building block for the creation of social entities as explained above.[7]

Human beings, along with many other animal species, are capable of engaging in cooperative behaviour and sharing attitudes among conspecifics (Searle, 2006, p. 16). Collective intentionality consists in (I) engaging in cooperative behaviour, (II) sharing intentional states, such as beliefs, desires, intentions (Searle, 1995, p. 152), or—as he rephrases it in his later analysis (2010, p. 43)—collective intentionality consists in (I.a) having collective distal intentions in cooperative planning (what he calls collective prior-intentions) and acting (what he calls collective intentions-in-action), and (II.a) holding collective distal intentions in believing and desiring. A complementary outline of the desiderata for the emergence of collective intentionality can be found in (Searle, 2010, pp. 44-45).

Rakoczy & Tomasello's (2007) criticise Searle for failing to appreciate the distinction between two different forms of group behaviour: social coordination and collective intentionality.[8] Rakoczy & Tomasello say that social coordination can be found in nonhuman animals' activities, while collective intentionality is human-specific. In order to justify this distinction, they provide evidence from empirical work on human infant development, and on the way children engage in social activities at different stages of their growth. Rakoczy & Tomasello emphasise the difference between the behaviour of chimpanzees and human children; they argue that Searle's claim that collective intentionality can be found not only in human behaviour, but also in the group activities of nonhuman animals—most notably in that of our nearest primate relatives—is wrong. Rakoczy & Tomasello say that during typical ontogeny, human children move from specific social interactions involving shared intentionality (learning, sharing, informing, helping) to participation in institutional contexts. They argue that if Searle is right and collective intentionality must exist inside individuals' heads, then the capacity for the creation of social entities requires the mastery of very sophisticated cognitive tools. They allude to the capacity for conceptual thinking (Tomasello likewise argued for it in 2014). Their point is that it is cognitively very demanding to grasp the content of the mental states of others when this content is conceptual and articulated through a symbolic system of reference, namely that of language (Rakoczy & Tomasello, 2007, p. 127). However, they infer a lack of cognitive capacities of chimpanzees just from the lack of motivational attitudes towards the accomplishment of certain group activities.

---

[7] Searle refers to "social facts" or "objective facts" rather than "social entities". But his use of the notion "social facts" is less pervasive than the use I would make of it, i.e., in Searle's terminology institutional reality is constituted by "social facts". He argues that all institutional facts are social facts, but not all social facts are institutional facts (Searle, 2014). Since I am primarily concerned with the constitutive elements of pre-institutional social reality, I avoid any terminological ambiguity and refer to rudimentary forms of social ontology as "social entities".

[8] In the comparative psychology literature, this latter notion is often used interchangeably with that of Shared Intentionality (see, especially, Tomasello & Carpenter, 2007; Tomasello et al., 2012; Tomasello, 2014).

Rakoczy and Tomasello do not claim that, necessarily, without language there is no collective intentionality, but rather that, necessarily, without those pre-linguistic social activities—such as role reversal imitation, which can be observed very early in human ontogeny (Tomasello & Call, 1997; Tomasello, 2003)—that constitute the normal route to the acquisition of a linguistic communicative modality, there is no collective intentionality. It follows that nonhuman group activities are classified by Rakoczy & Tomasello as forms of social coordination. But, as argued below, they are too quick in labelling these behaviours as forms of "complex social coordination only" (Rakoczy & Tomasello, 2007, p.116), especially when referring to what they admit to be one of the most sophisticated instances of group behaviour in the animal kingdom, i.e. chimpanzees' group hunting (see also, Tomasello, 2014, p. 35).

A subsequent study reinforces the above point that the limitations that Tomasello and colleagues worry about are more likely to be motivational than cognitive. Melis & Tomasello (2013) present the results of the first experimental study showing chimpanzees manifesting collaborative tendencies when engaging in the same coordinated action. Chimpanzees display attention to the partner, and some form of knowledge about the role played by the partner. Melis & Tomasello are cautious about over-interpreting these findings. But, at the very least, this study shows that chimpanzees have some appreciation of the fact that the presence of an interacting partner is related to their own goal, and, crucially, that the engagement of the partner will determine the success or failure of the intended action (they also observe a number of limitations that I do not discuss here; they are analysed at length in Melis & Tomasello, 2013). In addition, a study by Krupenye et al. (2016) shows that great apes can operate, at least at the implicit level, with an understanding of the behaviour of their conspecifics, manifestly displayed by the anticipation of the goal-directed actions of others.

It is important that the limitations in the collaborative performances of the chimpanzees tested by Melis & Tomasello might well be more motivational than cognitive, because, it implies that if chimpanzees are choosing not to cooperate, this is not due to a lack of understanding on their side, but to a lack of interest in the offer. This means that, if they have the choice, chimpanzees prefer to act alone. But, if needed, they interact collaboratively. Thus, the *can* question becomes a *want* question. The motivational question is a fascinating one, but not the one addressed here, and not needed to defend the evolutionary primitiveness of collective intentionality.

I argue below that "hunting teams" count as a by-product of chimpanzees' rich understanding of the social nature of their joint achievements. But I recognize that it is, indeed, very controversial to what degree chimpanzee hunting (or any other form of cooperation) depends on a) coordinated and intentional collaboration or b) a simpler mechanism such as coordinated actions in which multiple individuals pursue the same goal at the same time but with different intentions.

One way of approaching the debate is by asking: what is the nature[9] of the content of the mental states that are required in collective intentional behaviour, i.e., the future-directed intentions required in structuring and coordinating action, which Bratman calls "plan-states" (Bratman, 2012, 2014) and I define as "distal intentions" to emphasize their functional role as mental causal antecedents of action planning. To illustrate, consider an example of a distal intention involved in collective intentions: a chimpanzee, before engaging in a group hunt, might anticipate the role it will play in driving prey toward another member of the group. This kind of forward-directed intention plays a key role in structuring the cooperative activity, making it more than just simultaneous but interdependent actions.

---

[9] Gallotti (2012) has a richer proposal. He argues that the debate on the primitiveness of social ontology is both metaphysical and epistemological.

Importantly, distal intentions, in the context of collective coordinated behaviour, are not merely "involved" in collective intentionality; they are required for structuring and sustaining coordinated action. Without distal intentions, individuals might act simultaneously toward a shared goal but without the necessary framework that maintains cohesion in their collaboration. Distal intentions function as critical structuring elements that ensure complementarity among participants' roles, rather than mere parallel activity.

Another way to approach the debate is by asking what relationship the content of these and other types of mental states have with linguistic capacities. One reason to support this view is that distal intentions often involve deliberation about future actions, requiring a cognitive framework that structures how an individual foresees and organizes steps toward achieving a goal. For example, when one decides in the morning to cook dinner in the evening, their distal intention involves conceptual representations of food, cooking methods, and timing. This type of structured foresight has led some scholars to argue that distal intentions necessarily involve conceptual content Allen & Bekoff, 1997; Audi, 1973; Beardsley, 1978; Bratman, 1987; Churchland, 1970; Davidson, 1980; Goldman, 1970; Searle, 1983).

Call this view *Conceptualism about Distal Intentions (CDI)*. Different approaches to this view share these assumptions:

1. All action plans include distal intentions.

2. All distal intentions include conceptual content.

3. All action plans include conceptual content.

However, these premises need further clarification. Not all of the authors cited explicitly discuss "distal intentions" or "action plans" in these terms. Instead, their work concerns the broader idea that future-directed intentions play a structural role in action planning and coordination. Here, "action plans" refer to structured sequences of intended behaviours aimed at achieving future goals. In cognitive and philosophical literature, action plans are often considered mental representations of steps toward goal achievement, a perspective consistent with Bratman's and others' discussions of planning agency. Clarifying this notion helps to show why CDI proponents argue that action plans necessarily contain distal intentions and why they regard conceptual content as integral to such planning.

CDI comes in two forms: call them *Anthropocentric Conceptualism (AC)* and *Non-Anthropocentric Conceptualism (NAC)*. AC says that the only cognitive systems whose mental states have conceptual content are those cognitive systems who have linguistic capacities, i.e., full-blown adult human beings (Davidson, 1980, 1984; Bratman, 1987). NAC says that nonlinguistic cognitive systems also have mental states whose content is conceptual (Searle, 1994; Allen & Bekoff, 1997; Beck, 2012, for an overview of theories of concepts; Machery, 2009, and Weiskopf, 2008, for a description of a psychological theory of concepts, and Allen, 1999, and McAninch, Goodrich, and Allen, 2009, for how animals might be said to have concepts).

Drawing on this distinction between two variants of the traditional way of conceiving the content of mental states, the relevant question is whether linguistic capacities are required in order to hold and share distal intentions. The distinction between AC and NAC hinges on the premise that all distal intentions have conceptual content, but they diverge on the role of linguistic capacity in conceptualization:

AC: Conceptual capacities and linguistic capacities are interdependent. Since conceptual thought is tied to language, nonlinguistic animals cannot form distal intentions or collective intentions.

NAC: Conceptual capacities do not depend on language. Therefore, nonlinguistic creatures can still be planning agents, capable of forming distal intentions.

Searle argues that conceptual thinking does not depend on the mastery of language, allowing nonlinguistic creatures to form planning states and engage in collective intentionality. This positions him with NAC. However, earlier I suggested that primitive collective intentionality requires a notion of distal intentions that does not necessitate conceptual capacities. This requires rejecting both AC and NAC. To clarify, my position extends NAC by allowing for distal intentions structured by nonconceptual content, thereby making collective intentionality possible even in cases lacking full conceptual thought, as I shall explan later on.

Rakoczy & Tomasello (2007) argue against Searle's attribution of the capacity for the creation of social entities to nonhuman animals along the lines of the AC, that is, the idea that the capacity for conceptual thinking and the capacity for language are mutually dependent (Rakoczy & Tomasello, 2007, p. 123, 131). This means that conceptual capacities require language and vice versa. Their view conflicts with their own work showing that prelinguistic capacities underlie both language and conceptual thought. This creates ambiguity: can conceptual thought occur independently of linguistic structures, or is language merely a medium for expressing pre-existing conceptual capacities? To resolve this, I propose the following refinements:

If conceptual capacities require linguistic mastery, AC is upheld.

If conceptual capacities exist independently of language and later support linguistic development, NAC is upheld.

If conceptual capacities do not depend on linguistic mastery or full conceptualization, a nonconceptual framework for distal intentions should be explored.

Therefore, Rakoczy & Tomasello's rejection of the idea that nonlinguistic animals can create social entities suggests that they think it is a necessary condition for creatures to be capable of creating social entities that they have mastery of conceptual (and, perhaps thus, linguistic - altough Tomasello 2016 has a more refined approach that may keep concepts and language separate) capacities. The possession of conceptual capacities depends on complex representational skills, such as those for natural language. Hence, a characterization of nonconceptual mental content would be beneficial for an account of distal intentions, when these mental states belong to or are ascribed to nonlinguistic creatures (guidelines for such an account can be found in the work of Peacocke, 1992, 2014, 2016; Schellenberg, 2013; Crane, 2014). In the following section, I unpack what nonconceptual mental content might be.

I aim to further the investigation of the conceptual structure that underpins the development of Searle's theory of social reality because that would result in a more fruitful application of his philosophical analysis for empirical purposes, including those of Tomasello and colleagues' work. To understand the stakes of this debate, it is helpful to situate CDI, both in its AC and NAC variants, within two broader meta-level positions on the relationship between language and thinking as I shall explain next.

## 3. The primitiveness of collective intentionality: a defence

In this section, (a) I refine the distinction between conceptual and linguistic capacities, and (b) I argue that to defend the primitiveness of some forms of collective intentionality we need a notion of distal intentions, that does not require the mastery of conceptual capacities. Among the elements needed for the creation of social entities there is a fundamental capacity to form distal intentions and ascribe distal intentions to others. This capacity can, but does not *need* to lead, as Searle would argue, to the ascription of collective intentions (Searle, 1990), but, more fundamentally, it enables the ascription of individual intentions with meshing sub-plans (Bratman, 1992, 2014). While the ascription of collective intentions likely requires conceptual thinking, this may not be the case for the ascription of individual intentions with meshing sub-plans (see Pacherie, 2007 and 2011, for a similar claim).

To illustrate, consider as a case of collective intention that does not require conceptual content: a group of chimpanzees engaged in cooperative hunting. Each chimpanzee coordinates its movements with the others, anticipating the actions of their peers in ways that suggest shared goals. However, their coordination does not necessitate the kind of abstract conceptual understanding required for propositional judgments. By contrast, a case that does require conceptual content would be a formal agreement among humans to establish a constitution—such an agreement presupposes linguistic and conceptual capacities for explicit rule formation and adherence.

Searle maintains that conceptual thinking does not depend on the mastery of language, allowing nonlinguistic creatures to be planning agents capable of forming distal intentions and engaging in forms of collective intentionality.

Despite the usefulness of Searle's distinction, his dual-intention theory[10] seems to raise problems of its own, and indeed—following Pacherie—I maintain that Searle's characterization of distal intentions (or prior intentions, in his terminology) cannot fully account for the mental states of nonhuman animals (see also Pacherie, 2000). This necessitates a closer examination of the role of conceptual and nonconceptual content in intentional states.

---

[10] Some kinds of intention can guide proximal actions, and some other kinds of intention can guide action planning. A general distinction between these two kinds of intention is determined by the temporal location of the action that the intention is meant to guide. A narrower distinction is that while some intentions guide actions directly, other intentions guide plans for actions. As a result, the cognitive function of intentions that guide proximal actions is the monitoring and guidance of ongoing bodily movements (Brand, 1984); while the cognitive function of intentions that guide future actions is the monitoring and guidance of a plan.

I will now explain how the debate on mental content interfaces with that of the evolutionary primitiveness of collective intentionality.

We may call the position to the effect that thinking depends on language—and therefore that the only linguistic animals are human beings—the Strong Dependency View (SDV). This stance is notoriously defended by Donald Davidson (1984). On the other hand, we may call Weak Dependency V (WDV) the view to the effect that thinking does not depend on language, and therefore that nonlinguistic animals (i.e., nonhuman animals) only lack the capacity to form some mental states, namely those that depend on the mastery of linguistic capacities. Searle defends the latter, and, accordingly, his work addresses the following questions: can we decide which intentional states require language and which ones do not? And then, to whom can we ascribe such and such intentional states (Searle, 1994)?

The two 'dependency' views—SDV and WDV—offer meta-level positions on the relationship between language and thinking, with implications for CDI.

SDV: Thinking depends on language. This view implies that AC is the only viable form of CDI because nonlinguistic creatures lack the linguistic structures necessary for conceptual thought. Davidson (1984) defends this position, arguing that fine-grained discrimination among intentional states (e.g., beliefs, desires) requires a linguistically articulated concept of those states. According to Davidson, the truth or falsity of intentional states hinges on a metalinguistic capacity to evaluate their content, which is inherently tied to language.

WDV: Thinking does not depend on language. This position supports NAC by allowing that nonlinguistic creatures can engage in conceptual thinking. Searle (1994) argues that intentional states with conceptual content can exist independently of language. He contends that perception and action-oriented mental states, which have their own success conditions (e.g., fulfillment or unfulfillment), precede the development of propositional content and linguistic capacities. According to the SDV, defended by Davidson (1984), where there is no language it is impossible to make a fine-grained discrimination among intentional states, like beliefs or desires, or intentions. This is because this fine discrimination is about the content of the intentional states taken into account. The central point is that in order to discriminate between the truth or falsity of an intentional state one must have a linguistically articulated concept of that intentional state. This is a two-steps claim. We can agree with the first part that having an intentional state requires the capacity to discriminate conditions that satisfy the intentional state from those that do not (see Davidson, 1984, p. 171; Searle, 1994, p. 211). But Searle says that " 'truth' and 'false' are indeed *metalinguistic* predicates, but more fundamentally they are *meta intentional* predicates. They are used to assess success and failure of representations to achieve fit in the mind-to-world direction of fit, *of which statements and sentences are a special case*" (Searle, 1994, p. 212). According to Searle, the second part of this statement has a false premise. And this is because perception and action-oriented mental states are forms of intentionality that come prior to forms of intentionality that involve propositional contents, like beliefs and desires. The reason why intentions can be nonpropositional is this: while propositional attitudes like beliefs or desires need to be assessed for truth or falsity, nonpropositional intentional states like intentions can only be evaluated as fulfilled or as unfulfilled. For the latter, no boundaries are imposed by logic. In 2015 Searle argued that: "It is an old mistake to suppose that if animals can think, then they must be able to think that they are thinking. And it is an extension of that mistake to suppose that if animals have complex intentional structures in perception and action, then they must be able to think about the content of these complex intentional structures." (Searle, 2015, p. 5). This additional statement reinforces Searle's approach to the characterisation of the content of mental states.

The WDV can only be analysed if we know which aspects of intentional states require language. There are two such aspects: first, the conditions of satisfaction of the intentional state are linguistic; second, the mode of representing the conditions of satisfaction of the intentional state is linguistic (Searle, 1994, p. 213). Searle says: "It seems obvious to me that infants and animals that do not in any ordinary sense have a language or perform speech acts nonetheless have intentional states" (Searle, 1983, p. 5), and he argues that before there was language humans might have had beliefs and desires (Searle, 1982, p. 177). In Searle's view, the possession of intentional states with semantic, i.e., conceptual, content is independent of the possession of language.

These distinctions clarify the relationship between CDI and the dependency arguments. CDI pertains specifically to the content of distal intentions, examining whether conceptual capacities are required for these future-directed mental states. In contrast, SDV and WDV explore whether language is a prerequisite for conceptual thinking more broadly, providing the meta-framework within which CDI is situated. This meta-framework is preparatory to the proposal that I anticipated offering.

To recapitulate. The SDV implies that AC is the only viable form of CDI, as nonlinguistic creatures lack the linguistic structures necessary for conceptual thought. Davidson (1984) supports this view, arguing that fine-grained discrimination among intentional states (e.g., beliefs, desires) requires a linguistically articulated concept of those states. This is because evaluating the truth or falsity of intentional states hinges on a metalinguistic capacity to assess their content, which is inherently tied to language. The WDV, in contrast, allows for the possibility that nonlinguistic creatures can engage in conceptual thinking, though it does not necessarily entail NAC. Searle (1994) argues that some intentional states with conceptual content can exist independently of language. However, the claim that WDV directly supports NAC would be strong; rather, WDV permits the possibility that nonlinguistic creatures have conceptual thought. In practice, animal cognition research broadly supports WDV for terrestrial vertebrates and many other species, making it a widely accepted baseline assumption rather than a controversial claim.

A key issue with SDV is that it equates "thinking" with "conceptual thinking," an assumption that is neither self-evident nor universally accepted. While Davidson's view might be read as an extreme version of AC, it does not accurately reflect the positions of researchers such as Rakoczy & Tomasello (2007). These authors do not argue that nonhuman animals entirely lack intentionality; rather, their view suggests a graded approach to the relationship between conceptual capacities and linguistic capacities. My view clarifies this distinction by emphasizing that prelinguistic capacities form a basis for both conceptual thinking and linguistic abilities without necessitating strict interdependence. I defend a position called Capacity-Based Dependency View (CBDV). This view emphasizes the role of prelinguistic capacities in collective intentionality. According to the CBDV, language is necessary for the construction of those mental states whose content is conceptual—that is, mental states that consist in making judgments. However, these conceptual states are not required for all forms of intentionality, particularly those relevant to collective intentions of the sort discussed in this paper.

The most basic notion of conceptual thought applies to intentional states that represent some particular as belonging to some general kind or represent several particulars as standing in some relation (Gauker, 2002, p. 687). The reason why language users can be said to engage in conceptual thinking is because they can overtly *say* of some particulars that they stand in some relation to one another, and they can draw inferences from sayings of this kind (Gauker, 2007, p. 135; see also, Chater & Hayes, 1994, p. 237).

In the present context, I need to clarify how I am analysing the very notion of mental content. It can be characterised in terms of the conditions of satisfactions that it determines or in terms of those with which it is identified. Alternatively, content can also be characterised with respect to the attitudes that are taken towards that content. These specifications are implemented by a further distinction, that is, that of content as conceptual or nonconceptual. This latter distinction consists in content being identifiable with the information that a cognitive system receives from the world or as the way in which information is received from the world by the cognitive system (Schellenberg, 2013).

The conceptual/nonconceptual content debate is complex (see, for instance, Stalnaker, 2003; Byrne, 2004). I draw on (Bermùdez, 2007) and take conceptual mental content to be the informational content of the mental state of a subject that needs to master the concepts required to specify that content in order to be in that state; and I take nonconceptual mental content to be the informational content of the mental state of a subject who does not need to master the concepts required to specify that content in order to be in that state. More specifically, the distinction between conceptual and nonconceptual thought is roughly as follows: conceptual mental content represents a particular as belonging to some kind or of some particulars as standing in some relation. These contents are inferentially related; whereas nonconceptual mental content represents a "similarity judgement" among different particulars, where no inferential process needs to occur (Gauker, 2005, p. 289). To be more precise, conceptual thinking is, arguably, propositional and it works by means of disjunctive exclusion through the attribution of truth-values.

Hence, when distal intentions belong to or are ascribed to nonlinguistic creatures, we can invoke a nonpropositional-based explanation, because these mental states do not need to have a logical form. It should be noted that the claim that nonconceptual content results in nonpropositional format holds if we accept Davidson's SDV. But if we side with the defenders of the WDV, propositionalism could also be an option. Sinhababu (2015), who defends propositionalism, warns philosophers against assuming that a mental state having a propositional format entails that the mental state has conceptual content. He points out that "Philosophers should not follow linguistic evidence to a psychological theory that cannot deliver good psychological explanations" (Sinhababu, 2015, pp. 14). I propose that if propositional format can be dissociated from conceptual content, that is to say, if we can conceive of mental states with a propositional format and nonconceptual content, then propositionalism offers a way of overcoming the problem of how we can ever individuate animal mental content.

Given this generally applicable characterization of what nonconceptual content is, we can add, following Schellenberg (2013, P. 279), that mental content can be understood in nonconceptual terms if we allow it to provide discriminatory capacities that establish accuracy or inaccuracy conditions, and that guide us through the identification of particular instances of the world. Discriminatory capacities allow for the representation of the world as being a certain way. On the basis of what a creature represents the world to be, she can exercise a similarity judgement, and reallocate possibilities through the estimation of different options and the control over the selected one. Accordingly, she can mentally act in order to prepare for distal-oriented actions.

In conclusion, a nonconceptual account of mental content for distal intentions should be saving compositionality (needed to articulate a plan) without entailing propositionalism (the attribution of truth-values). Prelinguistic capacities play a foundational role in both language development and collective intentionality, serving as a common substrate that allows for the emergence of more complex cognitive and social structures. This claim differentiates the CBDV from both SDV, which ties fine-grained intentionality strictly to language, and WDV, which sometimes underplays the role of foundational capacities in enabling collective intentions. An example of how an advocate of WDA, such as Searle, has underplayed these foundational capacities can be found in his focus on language-independent conceptual thinking while paying less attention to the mechanisms that enable complex, structured cooperation in nonhuman animals - the sort discussed in the following section. For instance, while Searle acknowledges that animals can engage in cooperative actions, he does not emphasize how these actions rely on nonconceptual cognitive mechanisms such as behavioural synchronization, implicit communication, and sensitivity to social contingencies—capacities that enable structured but nonconceptual collective intentionality. CBDV highlights these mechanisms as crucial in bridging the gap between individual and collective intentionality.

A further clarification concerns how foundational capacities, sometimes called prelinguistic capacities, contribute to collective intentionality. These capacities include (1) the ability to form distal intentions, (2) the ability to attribute distal intentions to others, and (3) the ability to engage in coordinated, goal-directed joint actions without requiring explicit conceptual understanding. The latter, as said, encompasses behavioural synchronization, implicit communication, sensitivity to social contingencies, and shared action representation. These foundational capacities, while nonconceptual, enable individuals to navigate complex social interactions and engage in structured cooperation.

These requirements are a necessary basis for the framework that we can employ in order to explain how a future-oriented mental state can result from a nonlogical form of reasoning, leaving aside the fact the we do not have other compelling reasons to endorse the view that reasoning through causal representing must entail mental states with a propositional structure and a conceptual content (see, Camp, 2007, 2009; Clark, 2013; Gärdenfors, 2003; Gauker, 2007; Rescorla, 2009a, 2013; Schellenberg, 2013). If we take this explanatory route, we are no longer committed to the idea that, since nonhuman animals do not possess concepts, then nonhuman animals cannot represent the world in a veridical way.

In the articulation of my view I rely, in part, on Davidson's two-step argument. He argues that (1) thoughts are made of propositional attitudes and that (2) the content of every mental state is conceptual (Davidson, 1984, p. 103). I then apply the consequent analysis of the CBDV to the debate on the primitiveness of collective intentionality.

Rakoczy and Tomasello also seem to follow a Davidsonian strategy as their view suggests that they are arguing that conceptual capacities and linguistic capacities are interdependent (Rakoczy & Tomasello, 2007, p. 127). Perhaps it is Davidson's influence that prevents them from appreciating certain features of Searle's arguments for the primitiveness of collective intentionality. This is because Davidson does not just argue for (1) the interdependency between conceptual and linguistic capacities, but he also claims that (2) conceptual thinking is the only modality of thinking in general, including, a fortiori, thinking involved in planning. It seems that while Searle takes from Davidson only the second claim, Rakoczy & Tomasello take both the first and the second claim.

The CBDV defends Searle's selective adoption of Davidson's claim that conceptual thinking is not the only modality of thinking relevant to planning. However, it departs from Davidson and Rakoczy & Tomasello by emphasizing that prelinguistic capacities are foundational for both collective intentionality and language, without requiring a strict dependency between the two. This position accommodates evidence of nonconceptual distal intentions in nonlinguistic animals.

I argue for the first but not for the second claim: distal intentions involved in planning, in the creation of collective intentionality, and in the assignment of functions (not of *status* functions) are, plausibly, nonconceptual in content when they belong to nonlinguistic animals. I defend this claim by exploiting the general framework developed by Bratman (1987; 2014) in his planning theory, and then apply that framework to argue for primitive collective intentionality. Primitive social entities exist upon the ascription of nonconceptual distal intentions. Primitive social entities exist upon the ascription of nonconceptual distal intentions.

By defending the CBDV, I argue that while language is needed for the construction of conceptual mental states, it is not essential for all forms of intentionality, particularly those involved in planning and distal intentions. In the light of the foregoing, I now discuss the empirical evidence: chimpanzee group hunting. This involves the ascription of distal intentions, which are the basic building blocks for the creation of social entities. Group hunting involves distal intention ascription because it can be explained according to Bratman's planning theory which gives distal intention ascription a key role in the coordination and planning of joint activities.

One important clarification to be made in the interpretation provided in this paper of Rakoczy and Tomasello's account is whether they truly endorse the view that a propositional or highly conceptual understanding of the plan is necessary for collective intentionality. This point is inferred from their original arguments from 2007, and it warrants revisiting in light of Tomasello's more recent work where a similar idea resurfaces in his discussion of various aspects of action theory.

In *A Natural History of Human Morality* (2016) and *The Moral Psychology of Obligation* (2020), Tomasello refines his stance on collective intentionality, particularly emphasising a key element of Bratman's view—the commitment to mutual support. Tomasello (2016, 2020) argues that in order to uphold this mutual commitment, individuals must not only have a plan but also possess an understanding of the roles played by others within that plan. Crucially, Tomasello's revised position would suggest (to a reader) that this understanding need not be fully linguistic and may only be partially conceptual. What becomes essential is the participant's grasp of not just the distal intentions (long-term goals) of others, but also their proximal intentions—the immediate actions others are undertaking to achieve those goals.

This shift in focus highlights that successful cooperation hinges on more than just knowing the shared objective. For Tomasello (2020), it is equally vital for participants to understand how others are working toward that objective, allowing for mutual adjustments and support as needed. This refinement in Tomasello's theory suggests that a full propositional understanding may not be a strict requirement for collective intentionality, as participants can coordinate and support each other's efforts even with a partially conceptual grasp of the plan. However, any such reading of Tomasello's position would encounter some challenges in its extension. He argues that this mutual commitment to support, in turn, fosters a sense of fairness in distributing rewards among collaborators. This conclusion is problematic, as it rests on a contentious interpretation of Gilbert's (Gilbert, 1989, 2009) non-reductive and metaphysically primitive conception of collective intentions. Tomasello seems to borrow from Gilbert's framework to propose that fairness naturally follows from the mutual recognition of roles and contributions, but this leap raises questions. Gilbert's model is resistant to reduction, maintaining that collective intentions cannot be entirely reduced to individual intentions, a point Tomasello appears to gloss over. Critically, Tomasello's 2020 work has prompted responses from both Bratman (2020) and Gilbert (2020), who challenge these expansions of his theory. Their critiques underscore the complexities of applying the concept of mutual support to fairness in collaboration, particularly when viewed through the lens of Gilbert's more robust metaphysical framework (Bratman, 2014; Gilbert, 2020).

In sum (so far), the debate on the primitiveness of collective intentionality hinges on whether conceptual and linguistic capacities are necessary prerequisites for the formation of distal intentions that underpin joint actions. Searle's WDV asserts that non-linguistic creatures can form intentional states and engage in collective intentionality without the need for conceptual thought. This stands in contrast to Davidson's SDV, which argues that without language, animals cannot make fine-grained discriminations among their intentional states.

By supporting the CBDV, I contend that although language is necessary for constructing conceptual mental states, it is not indispensable for all types of intentionality, especially those related to planning and distal intentions. Non-linguistic animals, such as chimpanzees, can create distal intentions that are nonconceptual in nature, which still allow them to engage in collective actions.

In the context of primitive collective intentionality, these nonconceptual distal intentions provide the building blocks for the creation of social entities, as evidenced in chimpanzee group hunting. This framework, grounded in Bratman's planning theory, supports the notion that nonhuman animals are capable of forming social entities without the need for full-blown conceptual thought, thus reinforcing the view that collective intentionality is evolutionarily primitive. Further empirical investigation, such as the analysis of chimpanzee cooperation in various forms, provides evidence for these claims, challenging reductionist views like those of Rakoczy and Tomasello. By revisiting Tomasello's later work, we see an evolution of his stance, which now acknowledges that mutual understanding and coordination in joint actions may not require a fully propositional or conceptual understanding.

## 4. Hunting teams as social entities

According to Bratman's Planning Theory, distal intentions guide action planning. They provide control over the action and stability to the plan. On top of this, his view about shared agency is that interlocking distal intentions control and structure shared intentional activity. This, in a nutshell, is Bratman's constructivist approach to frame shared agency: "To intend to do something—Bratman argues—is to be in a plan state, where we understand what a plan state is by explaining its role in the rational dynamics of planning agency. Intending leads to action in ways that normally involve diachronic planning structures" (Bratman, 2012, p. 4).

Individual action planning is enabled by two features: firstly, the capacity for temporal extended intentional agency, which consists in the capacity to appreciate the place that one's own acting has within a broadly structured action, plus the acknowledgement that one's own activity is practically committed to that of the broadly structured action; secondly, the capacity for self-governance, which is the ability to take a practical standpoint as a guiding principle for one's own acting.[11] Bratman explains that in order to reach this further step, that is, shared intentional activity, we need to add to these two features a third one: the capacity to interlock individual distal intentions, and to ascribe distal intentions with meshing sub-plans. In sum: the capacity to create and ascribe individual distal intentions allows (1) extended intentional agency, (2) self-governance, and (3) shared intentional activity. These features are necessary and sufficient conditions for the emergence of collective intentionality.[12]

---

[11] See Holton, 2009, pp. 5-9 for different ways of understanding control and stability as features of intention; see also Shepherd, 2014, for a full account of control).

'Action-free perdurance' is introduced here as a concept to frame the proto-institutional aspect of hunting teams. This notion refers to the capacity of a social entity to exist as a normative structure solely during the performance of coordinated actions, without persisting beyond that timeframe. I argue that this proto-institutional capacity underpins the structure of hunting teams as primitive social entities.

Hunting practices among Ivory Coast chimpanzees can be framed in terms of Bratman's planning theory (Bratman, 1987). If Taï chimpanzees are capable of planning, as can be inferred from their behaviour, then they possess the capacity to form distal intentions. Since their actions are coordinated over time, according to Bratman's shared agency theory (Bratman, 2009c; 2014), Taï chimpanzees are also capable of sharing joint distal intentions. Taï chimpanzees' group hunting exhibits the three faculties analysed by Bratman. This behaviour can be labelled as a coordinated planning activity, and it also meets Searle's desiderata for any account of collective intentionality. Therefore, what I call *hunting teams* should be acknowledged as social entities.

While I argue that language facilitates the establishment and perdurance of institutional realities, the claim that normative states must be facilitated by language requires further empirical support. Recent research suggests that nonhuman animals, including chimpanzees, may possess forms of normative cognition (Andrews, 2020; Westra et al., 2024). These studies indicate that certain animals might operate within social norms that regulate behaviours within their communities, potentially challenging the assumption that normative states necessitate language. For example, studies on social normativity in primates suggest that chimpanzees engage in sanctioning behaviours when social expectations are violated (von Rohr et al., 2012). Similarly, bonobos display behaviours that indicate an awareness of social expectations and cooperative fairness (Tan, Ariely, and

---

[12] Philosophers already explored alternative approaches to the planning theory in order to make it suitable for explaining the behaviour of creatures that lack conceptual capacities (see Pacherie, 2007, 2011; Butterfill, 2011a). But none of these other proposals have taken into account the possibility to appreciate the planning theory in the light of a notion of nonconceptual mental content. Bratman has not overtly argued against any nonconceptual account of individual distal intentions with meshing sub-plans, nor has he denied the attribution of such distal intentions to nonhuman animals (Bratman, 2014, pp. 106, 185-186).

Hare, 2017). Some researchers even argue that social norms might be present in eusocial species, such as ants, where collective enforcement mechanisms regulate colony behaviour (Lorini, 2024). However, the nature of these norms and whether they require conceptual cognition remains contested. Tomasello (2020) argues that norms require shared intentions, which in turn demand abstract and conceptual capacities. While I acknowledge this perspective, my account suggests that normative regulation may exist on a spectrum, with some forms of social normativity emerging without the full conceptual apparatus required for institutional facts. If certain primates exhibit behaviours that function as social norms without relying on language, this challenges the strict dependency of normative states on conceptual thought.

Both the pygmy chimpanzee or bonobo (*Pan paniscus*) and the chimpanzee (*Pan Troglodytes*) cooperate in a wide range of situations and it has been observed that their cooperative behaviours are very plastic across a range of ecologies (see Muller & Mitani, 2005 and Hare & Tan, 2012, for a comprehensive review). For example, both bonobos and chimpanzees have long-term relationships in which they support each other through grooming, coalitions, and food sharing (de Waal, 1982, 1997; Nishida, 1983; Goodall, 1986; Kano, 1992; Parish, 1994; Vervaecke et al., 2000; Watts, 2002; Hohmann & Fruth, 2002). In addition, male chimpanzees have been observed to regularly hunt monkeys and patrol their territory borders in groups (Nishida, 1979; Wrangham, 1999; Boesch & Boesch-Achermann, 2000; Watts & Mitani, 2001; Mitani & Watts, 2001; Williams et al., 2004).

However, it is important to acknowledge that an alternative interpretation of this behaviour has been proposed by Moll and Tomasello (2007). They argue for a deflationary account of chimpanzee hunting, which denies the presence of shared goals or roles. According to their view, chimpanzees do not act with a sense of joint intentionality during group hunting but rather coordinate their behaviour individually without any true shared commitment to a collective goal. Tomasello's later work supports this interpretation, claiming that the kind of sharing required for collective intentionality is absent in the aftermath of chimpanzee hunting parties (Tomasello, 2020).

While Moll and Tomasello's deflationary account presents a legitimate challenge to the interpretation of chimpanzee hunting as a collective action involving shared goals, I contend that their argument may oversimplify the behaviours observed by Boesch and Boesch-Achermann. For instance, their critique primarily focuses on the absence of human-like social norms or fairness in the distribution of rewards after a hunt. However, the coordinated division of labour during the hunt, along with the stable roles each individual takes on, suggests a level of cooperation that cannot be easily reduced to mere individualistic action.

Moreover, recent research provides additional evidence for joint intentional action in nonhuman primates, which strengthens the case for interpreting chimpanzee hunting as involving collective intentionality.

Heesen et al. (2021) show that chimpanzees demonstrate joint commitment during cooperative activities. This suggests that they can coordinate their behaviour based on shared goals, fulfilling a core element of Bratman's planning theory. This joint commitment implies that chimpanzees are aware of not only their own role but also how their actions mesh with those of their partners. Duguid and Melis (2020) explore the mechanisms underlying animal collaboration, demonstrating that chimpanzees are aware of their partners' roles during problem-solving tasks. This understanding of proximal plans—where individuals adjust their behaviour in response to the immediate actions of others—shows a level of cooperation that supports the idea of shared intentionality in nonhuman animals. Additionally, Melis and Tomasello (2019) highlight how chimpanzees use communicative signals to coordinate their actions during joint problem-solving tasks. This communication is key to understanding how they maintain coordination and mutual responsiveness during group hunting, particularly as they switch between different roles such as drivers, blockers, chasers, and ambushers (Boesch & Boesch-Achermann, 2000). Buttelman et al. (2017) further support this by showing that great apes can distinguish between true and false beliefs during interactive tasks, revealing their capacity for understanding the intentions and mental states of others. Although their cognitive abilities may not be on par with human theory of mind, these findings indicate that chimpanzees possess a sophisticated social cognition that allows them to coordinate complex group actions.

These studies collectively bolster the argument that chimpanzees are capable of understanding both proximal and distal plans in ways that allow for more than coordination during collaborative tasks. In the remainder of the section, I re-analyse certain features of Taï Chimpanzees' group hunting (Boesch & Boesch-Achermann, 2000). This is, arguably, a case of ascription of individual distal intentions with meshing sub-plans that create a rudimentary social entity: a *hunting team*.

I take these cases of group hunting, studied by Boesch and colleagues, to provide evidence that nonlinguistic animals are capable of exercising collective intentionality. This is an instance of social reality because the social entity *hunting team* satisfies Searle's desiderata for an entity to count as social: that all intentionality, collective or individual, exists inside individuals' heads, and that individuals share collective goals and intend to do their own part in achieving the goal.

Accordingly, a hunting group is defined as such when: "individuals actively take part in a hunt by placing themselves in positions where they could perform a capture" (Boesch and Boesch-Achermann, 2000, p. 174)

Here is a brief summary of the group hunting practice. These strategies require a coalition of up to four individuals. Tai chimpanzees can succeed in coordinating a single hunt as long as each of the participants to the hunt remains loyal to his role. Four roles have been identified: the driver, the blocker, the chaser, and the ambusher. Roles are assigned on the basis of the current position of each chimpanzee with respect to the perceived location of the target. The reward is distributed proportionally to the role covered during the hunting. Individuals who worked harder and exposed themselves to higher risks get more than the others when it comes to sharing the spoils. In brief, this is the optimal hunting scenario: after locating the target, the driver forces the prey, i.e., normally a red colobus monkey, to take a specific direction in the canopy, and, at the same time, the blocker makes sure that the prey cannot deviate from the direction forced by the driver. Then, the chaser tries to catch the prey by climbing under it. Then, the ambusher smoothly blocks the escape and traps the red colobus.

Boesch and Boesch-Achermann (2000) say that cooperation in hunting is kept stable by: (a) a mechanism for individual recognition, (b) temporary memory of actions in the recent past, (c) attribution of value to those actions, and (d) social enforcement of those values. Notably, the features pointed out by Bratman's Planning Theory correspond to the features that keep the hunting stable. The capacity for temporally extended intentional agency depends on the ability to (b) retain memory of actions in the recent past. This supports the capacity for self-governance that depends on the ability to (c) attribute values to the actions in which the agents are engaged. The latter triggers (d) mutual responsiveness to those actions in virtue of the values that are attributed to them. Then, this results in (a) a mechanism for mutual recognition that enables shared intentional activity.

Blomberg (2015) offers a sympathetic analysis to my view. He highlights the difference between collective and distributive action, exploring how chimpanzees engage in what appears to be collective action during group hunting. This is contrasted with Tomasello's argument that such actions are individualistic. Blomberg critiques the dualism between true collaboration and mere coordination, arguing for a middle ground where joint action does not necessarily require full collaboration but involves coordinated actions with a common goal. I argue that nonhuman animals create social entities through collective intentionality, even if this intentionality is metaphysically reducible to individual intentions. This reduction aligns with Blomberg's middle-ground approach, suggesting that what may appear as individualistic behaviour can still form the basis of a social entity. Blomberg supports the idea that chimpanzees can coordinate their behaviour toward a common goal (e.g., capturing prey), even if their intentions remain individualistic. I have indirectly built on this by arguing that these individual intentions create social entities, such as hunting teams, without requiring the sophisticated cognitive capacities that human institutions may demand. We both touch upon the idea of non-conceptual content in joint actions. Blomberg's approach implies that collective goals can emerge without the need for agents to fully represent those goals conceptually. And the argument offered in this paper similarly emphasises that nonhuman animals' collective intentions do not necessitate conceptual thought, aligning with Searle's WDV.

On a similar note, evidence for collaborative hunting among nonhuman primates may even tell us about the similarities of the capacity for action planning built on joint intentions in species other than us. Interestingly, drawing on studies on hunter-gatherer human populations in Paraguay and Venezuela (Kaplan et al., 1985; Walker et al., 2002), the learning of hunting behaviour of humans has analogous developmental paths to those of Tai chimpanzees, both in terms of the time necessary for learning and of the age-range during which individuals hunt more frequently and more efficiently. As Boesch explains, hunting behaviour is a learning process that starts around 8–10 years and that takes about 20 years of practice in order to be mastered. Different roles in the hunt require different levels of expertise and can be performed by more or less experienced individuals. This practice is very demanding because it requires: 1) the capacity to understand the behaviour of another species, i.e., the red colobus monkeys, and 2) the capacity to coordinate actions among individuals towards a common goal across time. A *hunting team* is thus described: it is a coalition of individuals, normally up to four, that requires a strategic coordination of the action of each member towards the same target.

Hunting teams can be acknowledged as instances of social entities upon two out of the three primitives outlined by Searle: 1) the capacity for collective intentionality, and 2) the capacity for the assignment of functions. These are both thought to be sufficient conditions for the creation of social entities. Action-free perdurance is what I take to be the main difference between social entities created upon these two capacities, and social entities that include the third primitive too, i.e., 3) the capacity for the assignment of status functions, that is, for the creation of institutional reality. Nevertheless, I maintain that institutional facts, as opposed to proto-normative structures, presuppose the ability to establish and sustain a normative state of affairs beyond immediate action. This capacity is more robustly enabled by language and conceptual cognition. However, further empirical research is needed to determine the extent to which nonhuman animals can maintain normative structures independently of language.

I emphasize that the concept of "action-free perdurance" captures the unique proto-institutional feature of these hunting teams, where the social structure exists only during the performance of the coordinated action.

Boesch describes a parallelism between activities such as those undertaken by hunting chimpanzees and football players (Boesch, 2005, p. 629), and we may appreciate the similarities between the two cases in the light of the description of hunting practices that I analysed above: "Like a team of soccer players, individuals react opportunistically to the present situation while taking into account the shared goal of the team. Some players will rarely make a goal, like defenders and goalies, but the success of the team will critically depend upon their contribution. This is very reminiscent of group hunting in chimpanzees where synchronisation of different coordinated roles, role reversal, and performance of less successful roles favour the realisation of the joint goal". However, there are significant differences between these two examples, which clarify why hunting teams are best characterized as proto-institutions rather than full institutions.

Specifically, social entities with institutional components, like football teams or political parties, exist both prior and subsequent to the online coordination of planned actions. Their existence and normative structure are independent of the actual performance of joint action. This action-free perdurance is a hallmark of institutions. By contrast, hunting teams lack this feature: their normative structure and social ontological status exist only in an online mode, during the actual performance of a joint action. For example, while a football team remains identifiable and normatively structured even when not playing a match, a hunting team is constituted only during the act of hunting. This distinction stems from the cognitive capacities required to create and maintain institutional realities. Institutional facts presuppose the ability to establish and sustain a normative state of affairs even in the absence of immediate action, a capacity facilitated by language and conceptual thought. Lacking these capacities, creatures such as chimpanzees are unlikely to create entities with the full normative power and perdurance of institutions. Instead, their coordinated hunting behaviours exemplify proto-institutions—rudimentary forms that depend on the immediate context of joint action.

Social entities with an institutional component exist prior and subsequently to the online coordination of planned actions. Their existence is independent of the actual or online performance of a joint action. Instead, for hunting teams, it seems that the normative aspect of the social ontological status of such entities only exists in an online mode, that is during the actual performance of a joint action, as I argued that cognitive capacities that allow for conceptual thought may be requested for creating institutional reality, and for maintaining the normative power of institutions. Since language seems to facilitate the establishment and the perdurance of a state of affairs in the absence of online coordination of joint actions, it follows that creatures lacking conceptual capacities, presumably, lack those for the creation of institutional facts as well. This is a valuable option that would need further examination[13].

## 5. Conclusion

Group hunting informs us about how chimpanzees manipulate their world—i.e., the niche they occupy in a given environment (see Cheney and Seyfarth, 1990), and about how they create social entities such as hunting teams.

---

[13] Given the complexity of this issue, the discussion on perdurance and institutional facts may warrant a separate analysis. While this paper introduces the idea of action-free perdurance as a distinguishing feature of institutional versus proto-institutional structures, a more comprehensive study of the empirical literature on nonhuman normativity may strengthen the theoretical framework. I acknowledge that this aspect of the argument requires further refinement and empirical engagement.

As explained in Section 2, Searle's assumes that the foundational primitive for the creation of social entities, i.e., the capacity for collective intentionality, "is the psychological presupposition of all *social* reality" (Searle, 2006, p. 16), where any social entity is such in that it involves collective intentionality on the side of two or more, human or nonhuman, conscious agents. Rakoczy & Tomasello (2007) accuse Searle of over-ascribing cognitive capacities to nonhuman animals in such a way that leads Searle to mark the uniqueness of the human mind too far, i.e., at the capacity for the creation of institutional reality rather than at the capacity for collective intentionality.

As shown in Section 3, this objection to Searle can be justified only if Searle characterization of the content of mental states as conceptual is correct. But the replacement of Searle's notion of the mental content of distal intentions with a less cognitively demanding notion makes room for the analysis of a large body of empirical work that, in principle, is compatible with Searle's intuition about the primitiveness of the capacity for collective intentionality.

The optimal functioning of *hunting teams* may contain the first and foundational primitive of social reality, i.e., the capacity for collective intentionality, and it is intriguing to think about it as a fitting piece of the puzzle in the evolutionary history of social reality. Crucially, as argued in Section 3, among the elements needed for the creation of social entities are the capacities to create individual distal intentions and to ascribe individual distal intentions to others, which results in the appreciation of joint distal intentions. Searle advocates the idea that social entities are created upon the ascription of irreducibly collective distal intentions, but this is not always the case. For the creation of some social entities, all that is required is just the ascription of individual distal intentions with meshing sub-plans, as Bratman advocates; and, while the ascription of irreducibly collective distal intentions seems to require the capacity for conceptual thinking, this may not be the case for the ascription of individual distal intentions with meshing sub-plans. There is much evidence that (at least rudimentary) social entities exist among animal species other than humans, and, if this is the case, we urge a theory of social reality that could account for that. If the ascription of distal intentions is necessary for the creation of social entities, and if the former can occur in the absence of the uniquely human capacity for concepts and language, then nonhuman animals may be capable of creating (primitive) social entities by ascribing mental states, such as some distal intentions, that are nonconceptual.

**Declarations**

**Disclosure of potential conflicts of interest:** I declare no conflict of interests.
**Funding:** Not applicable.
**Ethical approval and Consent to Participate:** Not applicable.
**Authors' contributions:** I am the only contributor.
**Data availability:** Not applicable.

Forthcoming in *Philosophia*

...